\documentclass[aps,prd,twocolumn,groupedaddress,floatfix]{revtex4-1}
\usepackage[dvips]{graphicx}
\usepackage{amsmath,amssymb}
\usepackage{hyperref}
\usepackage{bm}
\usepackage{longtable}
\usepackage{numprint}

\newcommand{\uvec}[1]{\bm{\hat{#1}}}

\begin{document}

%\preprint{}

\title{Spin effects in the phasing of gravitational waves from binaries on
eccentric orbits}


\author{Antoine Klein}
\email{aklein@physik.uzh.ch}
%\thanks{}
\affiliation{Institut f\"ur Theoretische Physik, Universit\"at Z\"urich,
Winterthurerstrasse 190, 8057 Z\"urich}

\author{Philippe Jetzer}
\affiliation{Institut f\"ur Theoretische Physik, Universit\"at Z\"urich,
Winterthurerstrasse 190, 8057 Z\"urich}

\date{\today}

\begin{abstract}
We compute here the spin-orbit and spin-spin couplings needed for an accurate
computation of the
phasing of gravitational waves emitted by comparable-mass binaries on
eccentric orbits at the second post-Newtonian (PN) order. We use a
quasi-Keplerian parametrization of the orbit free of divergencies in the zero
eccentricity limit. We find that spin-spin
couplings induce a residual eccentricity for coalescing binaries at 2PN,
of the order of $10^{-4}$-$10^{-3}$ for supermassive black hole binaries in
the LISA band. Spin-orbit precession also induces a non-trivial pattern in the
evolution of the eccentricity, which could help to reduce the
errors on the determination of the eccentricity and spins in a gravitational
wave
measurement.
\end{abstract}

%\keywords{}

\maketitle

\section{Introduction}

Up to now most studies concerning gravitational wave emission from binaries
have been done assuming circular orbits (see e.g.~\cite{petf,kjs}
and references
therein). Numerical studies of the formation
of binaries and of their subsequent development suggest instead
that the orbits could be eccentric~\cite{caab,bpbms}. It is thus of relevance to
take
the eccentricity
into account when one computes the waveform emitted by such a system.
This problem has not yet been fully explored, mainly due to its 
complexity, and here as a further step we derive the equations 
which govern the time evolution of the orbital parameters,
and in particular the eccentricity,
including the spin-orbit and the spin-spin couplings needed for an accurate
computation in the post-Newtonian approximation up to 2PN of the phasing of the
gravitational waves emitted
by binary systems, with components of comparable mass.

The PN approximation is
valid when the two object forming the binary are sufficiently separated. The
issue of estimating the limit of its validity has been tackled with different
methods:
comparisons
between
post-Newtonian templates and results from numerical relativity, for non-spinning
binaries on quasi-circular
orbits~\cite{bcgshhb}; and comparisons between different post-Newtonian
waveforms for spinning and non-spinning binaries on quasi-circular
orbits~\cite{bcv1,bcv2}.
All of these studies found a remarkable
reliability of the PN approximation up to separations as small as the innermost
stable circular
orbit, $r=6GM/c^2$. It is not clear if that still holds for eccentric binaries,
and this
must be answered by extending these comparisons to such systems, but one can
trust that for low enough eccentricities, the post-Newtonian
approximation is reliable up to the end of the inspiral.

We also derive a quasi-Keplerian parametrization of the orbit free of
divergencies in the zero eccentricity limit, and find that spin-spin couplings
induce a residual eccentricity of 2PN order after the orbit has been
circularized by gravitational wave emission.

We then solve the equations which govern the evolution of the eccentricity
and the mean motion for different values of the masses
and spins as a function of the initial eccentricity. 

\section{Kepler equations and evolution of the mean motion and of
the eccentricity}

As spin-orbit couplings appear at 1.5PN order and spin-spin couplings at 2PN
order, it is sufficient to consider
only the Newtonian and spin-coupling term in the equations of motion. For
simplicity, we will use a system of units where $G=c=M=1$, where $M$ is the
total mass of the system.
We start from the generalized Lagrangian in the center of mass frame used
in~\cite{kww,kidder}:
\begin{align}
 \mathcal{L} &= \frac{\nu}{2} \bm{v}^2 +\frac{\nu}{r} +\frac{\nu}{2} (\bm{v}
\times
\bm{a} ) \cdot \bm{\xi} - \frac{2\nu}{r^3} (\bm{x} \times \bm{v}) \cdot
(\bm{\zeta} + \bm{\xi}) \nonumber\\
&+ \frac{1}{r^3} \bm{S}_1 \cdot \bm{S}_2 - \frac{3}{r^5} \left( \bm{x} \cdot
\bm{S}_1 \right) \left( \bm{x} \cdot
\bm{S}_2 \right) ,
\end{align}
where
\begin{align}
 \nu &= m_1 m_2, \\
 r &= |\bm{x}|, \\
 \bm{\zeta} &= \bm{S}_1 + \bm{S}_2, \\
 \bm{\xi} &= \frac{m_2}{m_1} \bm{S}_1 + \frac{m_1}{m_2} \bm{S}_2 .
 \end{align}

The equations of motion are
\begin{align}
 p^i &= \frac{\partial \mathcal{L}}{\partial v^i} - \frac{d}{dt} s^i, \\
 \frac{dp^i}{dt} &= \frac{\partial \mathcal{L}}{\partial x^i},
\end{align}
where $s^i = \partial
\mathcal{L}/\partial a^i$.

We can solve them order by order, which gives at 2PN order
\begin{align}
 \bm{p} &= \nu \bm{v} + \frac{\nu}{r^3} \bm{x} \times (2 \bm{\zeta} + \bm{\xi}),
\end{align}
\begin{align}
 \bm{a} &= - \frac{\bm{x}}{r^3} + \frac{\bm{x}\cdot \bm{v}}{r^5} \bm{x} \times
(6\bm{\zeta} + 3 \bm{\xi}) \nonumber\\
&- \frac{1}{r^3} \bm{v} \times (4
\bm{\zeta} + 3 \bm{\xi}) + \frac{\bm{x}}{r^5} (\bm{x} \times \bm{v} ) \cdot
(6\bm{\zeta} + 6\bm{\xi}) \nonumber\\
&- \frac{3\bm{x}}{\nu r^5} \bm{S}_1 \cdot \bm{S}_2- \frac{3}{\nu r^5} \left[
\left(
\bm{x} \cdot \bm{S}_2 \right)
\bm{S}_1 + \left( \bm{x} \cdot \bm{S}_1 \right) \bm{S}_2 \right]  \nonumber\\
& + \frac{15\bm{x}}{\nu r^7} \left(
\bm{x} \cdot
\bm{S}_1 \right) \left( \bm{x} \cdot
\bm{S}_2 \right).
\end{align}

The reduced energy and reduced orbital angular momentum are given by
\begin{align}
 \bm{J} &= \frac{1}{\nu} \left( \bm{x} \times \bm{p} + \bm{v} \times \bm{s}
\right) \nonumber\\
 &= \bm{x} \times \bm{v} + \frac{1}{r^3} \bm{x} \times [ \bm{x} \times (2
\bm{\zeta} + \bm{\xi} )] - \frac{1}{2} \bm{v} \times (\bm{v} \times \bm{\xi}),
\label{Jofxv}
\\
  E &= \frac{1}{\nu} \left( \bm{p} \cdot \bm{v} + \bm{s} \cdot \bm{a} -
\mathcal{L} \right) \nonumber\\
 &= \frac{1}{2} \bm{v}^2 - \frac{1}{r} + \frac{1}{r^3} (\bm{x} \times
\bm{v})
\cdot \bm{\xi} \nonumber\\
&- \frac{1}{\nu r^3} \bm{S}_1 \cdot \bm{S}_2 +
\frac{3}{\nu r^5} \left( \bm{x} \cdot
\bm{S}_1 \right) \left( \bm{x} \cdot
\bm{S}_2 \right). \label{Eofxv} 
\end{align}

The magnitude of $\bm{J}$ is not constant along an orbit~\cite{gergely}. Indeed,
due to spin-spin interactions, both spin
vectors undergo a precessional motion and thus, from the conservation of
the total angular momentum, it follows that $J$ changes at the 2PN order.
If we denote its angular average (with respect to the true anomaly $v$,
defined later) by
$L$, and define $A = \sqrt{1 + 2EL^2}$, we get
\begin{align}
 J &= L - \frac{1}{2\nu L^3} \left| \uvec{J} \times \bm{S}_1 \right| \left|
\uvec{J}
\times \bm{S}_2 \right| \{2A \cos(v-2\psi) \nonumber\\
&+ (3 + 2A \cos v )\cos[2(v -
\psi)] \} = \nonumber\\
&= L - \frac{\gamma_2}{2L^3} \{ 3A \cos(v-2\psi) \nonumber\\
&+ 3 \cos[2(v - \psi)] +
A \cos(3v - 2\psi)\}, \\
 \gamma_2 &= \frac{1}{\nu} \left| \uvec{J} \times \bm{S}_1 \right| \left|
\uvec{J}
\times \bm{S}_2 \right|,
\end{align}
where we defined $\psi$ to be the angle subtended by the bisector of
the projections of $\bm{S}_i$ in the plane of motion and the periastron line.

We can find a quasi-Keplerian solution to these equations, as (see the
appendix)
\begin{align}
 r &= a \left( 1 - e_r \cos u \right) + f_r \cos[2(v - \psi)],
\label{Keplereqradius} \\
 \phi &= (1+k) v + f_{\phi,1} \sin(v-2\psi) + f_{\phi,2} \sin[2(v-\psi)],
\label{phiofv}\\
 v &= 2 \arctan \left( \sqrt{\frac{1 + e_\phi}{1 - e_\phi}} \tan \frac{u}{2}
\right), \\
 l &= n(t-t_0) = u - e_t \sin u,
\label{Keplereqtime}
\end{align}
where $(r,\phi)$ is a polar coordinate system in the plane of motion, $n$ is the
mean motion, $u$, $v$, and $l$ are the eccentric, true, and mean anomalies, $a$
is
the
semi-major
axis, $e_t$, $e_r$, and $e_\phi$ are eccentricities, $k$ accounts for
perihelion precession, and the $f_i$ are constants.

This parametrization is different from the one found in~\cite{kmg}, which
suffered from an apparent singularity in the limit $e \to 0$ (all three
eccentricities tend together towards zero). This singularity was due to the
fact that the authors used as a definition of the eccentric anomaly, denoted in
their paper by $\xi$,
\begin{equation}
 r(\xi) = \frac{1}{2} \left[ r_{\mbox{max}} + r_{\mbox{min}} - \left(
r_{\mbox{max}} -  r_{\mbox{min}} \right) \cos\xi 
\right],
\end{equation}
which leads to Eq.~\eqref{Keplereqradius} with $f_r = 0$. The zero
eccentricity limit of the equations of motion $r(\phi)$ and $\dot{\phi}$ leads
to $r = \bar{r} + \delta r \cos[2(\phi-\psi)]$ (see the appendix). If $f_r = 0$
in
Eq.~\eqref{Keplereqradius}, this angular dependence must come from the 
change of variables $u(\phi)$. To cancel the $e_r = O(e)$ factor in front of
$\cos(u)$ so that the angular
dependence does not vanish in the zero eccentricity limit, the function
$u(\phi)$ must be of order $O(e^{-1})$. This is the origin of the apparent
singularity in the quasi-Keplerian parametrization found in~\cite{kmg}.

Our parametrization has the advantage of being free from singularities in the
zero eccenticity limit, so that the latter can be more transparently studied.
Note, however, that the periastron line (defined by the equation $u = v
= 2 p \pi$, $p \in \mathbb{Z}$) does no longer correspond to $r =
r_{\mbox{min}}$.

The mean motion and time eccentricity are given, in terms of $E$ and
$L$, as
\begin{align}
 n &= (-2 E)^{3/2}, \label{nofEL} \\
 e_t^2 &= A^2 + \frac{E}{L} \beta\left(8 , 6 - 2A^2 \right)  + 2 \frac{E}{L^2}
\gamma_1,
\label{eofEL}
\end{align}
where
\begin{align}
 \beta(a,b) &= \uvec{J} \cdot \left( a \bm{\zeta} + b \bm{\xi} \right), \\
 \gamma_1 &= \frac{1}{\nu} \left[ \bm{S}_1 \cdot \bm{S}_2 - 3 \left( \uvec{J}
\cdot \bm{S}_1 \right)
\left( \uvec{J} \cdot \bm{S}_2 \right) \right].
\end{align}

We can invert these relations and find $E$ and $L$ as functions of the
post-Newtonian parameter $x = n^{2/3}$ and the eccentricity $e = e_t$. These are
\begin{align}
 E &= - \frac{x}{2}, \label{Eofex} \\
 L &= \frac{\sqrt{1 - e^2}}{x^{1/2}} \left[ 1 - \frac{x^{3/2} \beta \left(4, 3-
e^2 \right)}{2\left(1-e^2\right)^{3/2}} - \frac{x^2
\gamma_1}{2\left(1-e^2\right)^2} \right]. \label{Lofex}
\end{align}

These allow us to express the constant parameters of the quasi-Keplerian
motion as
\begin{align}
 a &= x^{-1} \left[ 1 + \frac{x^{3/2} \beta(2,1)}{\sqrt{1-e^2}} +
\frac{x^2 \gamma_1}{2\left(1-e^2\right)} \right] ,
\label{aofex}
\end{align}

\begin{align}
 k &= -\frac{x^{3/2} \beta\left(4,3\right)}{\left(1-e^2\right)^{3/2}} -
\frac{3x^2
\gamma_1}{2\left(1-e^2\right)^2} , \label{kofex} \\
 e_r &= e \left[ 1 - \frac{x^{3/2} \beta \left(2, 1
\right)}{\sqrt{1-e^2}} - \frac{x^2 \gamma_1}{2\left(1-e^2\right)} \right] ,
\label{erofex}
\\
 e_\phi &= e \left[ 1 - \frac{x^{3/2} \beta \left(2, 2
\right)}{\sqrt{1-e^2}} - \frac{x^2 \gamma_1}{\left(1-e^2\right)} \right]
\label{ephofex},\\
f_r &= - \frac{x}{2\left(1 - e^2\right)} \gamma_2,\\
f_{\phi,1} &= - \frac{e x^2}{\left(1 - e^2\right)^2} \gamma_2,\\
f_{\phi,2} &= - \frac{x^2}{4 \left(1 - e^2\right)^2} \gamma_2.
\label{fphi3ofex}
\end{align}

We can now use the results from~\cite{gpv,gergely}, where the
orbit averages of $dE/dt$ and $dL/dt$ due to the emission of gravitational waves
were computed:
\begin{align}
 \frac{dE}{dt} &= \nu \left(\dot{E}_N + \dot{E}_{SO} + \dot{E}_{SS} \right),
\label{Edot}\\
\frac{dL}{dt} &= \nu \left(\dot{L}_N +
\dot{L}_{SO} + \dot{L}_{SS} \right), \label{Ldot}
\end{align}
where
\begin{widetext}
\begin{align}
 \dot{E}_N &= - \frac{(-2E)^{3/2}}{15 L^7} \left( 96 + 292 A^2 + 37 A^4
\right), \\
 \dot{E}_{SO} &= \frac{(-2E)^{3/2}}{10 L^{10}} \beta \left( 2704 + 7320A^2
+2490 A^4 + 65 A^6 , 1976 + 5096A^2 + 1569 A^4 + 32 A^6
\right), \\
\dot{E}_{SS} &= \frac{(-2E)^{3/2}}{960 L^{11}} \big[ 2
\sigma\big( \numprint{42048} + \numprint{154272}A^2 + \numprint{75528} A^4 +
3084 A^6, \numprint{124864} + \numprint{450656}A^2 + \numprint{215544} A^4 +
8532 A^6 , \nonumber\\
& \numprint{131344}A^2 + \numprint{127888} A^4 +
7593 A^6 \big) - \tau\big( 448 + 4256 A^2 + 3864 A^4 + 252 A^6 ,64 + 608 A^2
+ 552 A^4 + 36 A^6, \nonumber\\
& 16 A^2
+ 80 A^4 + 9 A^6 \big)
\big] ,
\end{align}

\begin{align}
 \dot{L}_N &= - \frac{4(-2E)^{3/2}}{5 L^4} \left( 8 + 7 A^2 \right), \\
 \dot{L}_{SO} &= \frac{(-2E)^{3/2}}{15 L^7} \beta \left( 2264 + 2784 A^2 + 297
A^4 , 1620 + 1852 A^2 + 193
A^4 \right), \\
 \dot{L}_{SS} &= \frac{(-2E)^{3/2}}{20 L^{8}} \big[2 \sigma\left(  552 + 996 A^2
+ 132 A^4 ,1616 + 2868
A^2 + 381 A^4 , 894 A^2 + 186 A^4 \right) \nonumber\\
&- \left( 8 + 24 A^2 + 3 A^4 \right)\tau\left( 2, 1 ,0 \right)
\big],
\end{align}

\begin{align}
 \sigma(a,b,c) &= \frac{1}{\nu} \left[ a \bm{S}_1 \cdot \bm{S}_2 - b \left(
\uvec{J}
\cdot \bm{S}_1 \right) \left( \uvec{J}
\cdot \bm{S}_2 \right) + c \left|
\uvec{J} \times
\bm{S}_1 \right| \left| \uvec{J}
\times \bm{S}_2 \right| \cos2\psi \right], \label{eqsigma}\\
\tau(a,b,c) &= \sum_{i=1}^2 \frac{1}{m_i^2}  \left[
 a \bm{S}_i^2 - b \left( \uvec{J} \cdot \bm{S}_i \right)^2 + c \left| \uvec{J}
\times
\bm{S}_i \right|^2  \cos 2\psi_i \right], \label{eqtau}
\end{align}
where $\psi_i$ is the angle subtended by
the projection of $\bm{S}_i$ in the plane of motion and the periastron line.

We can express these orbit averages
in terms of $x$ and $e$ using the post-Newtonian expressions~\eqref{Eofex}
and~\eqref{Lofex}. Using Eqs.~\eqref{nofEL} and~\eqref{eofEL}, we find the
time derivatives of the mean motion and of the eccentricity:
\begin{align}
 \frac{dn}{dt} &= \frac{\nu x^{11/2}}{\left(1 - e^2\right)^{7/2}} \Bigg[
\frac{1}{5} \left( 96 + 292 e^2 + 37 e^4 \right) \nonumber\\
&-
\frac{x^{3/2}}{10\left(1 - e^2\right)^{3/2}} \beta \left( 3088 +
\numprint{15528}e^2 + 7026 e^4 + 195e^6, 2160 + \numprint{11720} e^2 + 5982 e^4
+ 207 e^6 \right) \nonumber\\
& - \frac{x^2}{160\left(1-e^2\right)^2} \sigma \big( \numprint{21952} +
\numprint{128544} e^2 + \numprint{73752} e^4 + 3084 e^6, \numprint{64576} +
\numprint{373472} e^2 + \numprint{210216} e^4 + 8532 e^6, \nonumber\\
& \numprint{131344} e^2 + \numprint{127888} e^4 + 7593 e^6 \big) \nonumber\\
&+
\frac{x^2}{320\left(1-e^2\right)^2} \tau \left( 448 + 4256 e^2 + 3864 e^4 + 252
e^6, 64 + 608 e^2 + 552 e^4 + 36
e^6, 16 e^2 + 80 e^4 + 9 e^6 \right) \Bigg], \label{ndot}
\end{align}

\begin{align}
 \frac{de^2}{dt} &= -\frac{\nu x^4}{\left(1 - e^2\right)^{5/2}} \Bigg[
\frac{2e^2}{15} \left( 304 + 121 e^2 \right) - \frac{e^2 x^{3/2}}{15\left(1 -
e^2\right)^{3/2}} \beta \left( \numprint{13048} + \numprint{12000} e^2 + 789
e^4 , 9208 + \numprint{10026} e^2 + 835 e^4 \right) \nonumber\\
&  - \frac{x^2}{240\left(1-e^2\right)^2} \sigma \big( -320 + \numprint{101664}
e^2 + \numprint{116568} e^4 + 9420 e^6, - 320 + \numprint{296672}
e^2 + \numprint{333624} e^4 + \numprint{26820} e^6,  \nonumber\\
& \numprint{88432}
e^2 + \numprint{161872} e^4 + \numprint{16521} e^6 \big) \nonumber\\
&+\frac{x^2}{480\left(1-e^2\right)^2} \tau \big( - 320 + 2720 e^2 + 5880 e^4 +
540
e^6, - 320 - 160 e^2 + 1560 e^4 + 180
e^6, 16 e^2 + 80 e^4 + 9 e^6 \big) \Bigg]. \label{edot}
\end{align}
\end{widetext}

We find perfect agreement with~\cite{gpv}, where the spin-orbit effects were
computed in terms of $a$ and $e_r$. One can worry that these derivatives depend
on the angles $\psi_i$, which are not well-defined in the circular limit.
This is however not a problem, as this dependence disappears in this limit both
for $dn/dt$ and $de^2/dt$.

We can see that the spin-spin couplings computed here induce a positive
derivative
$de^2/dt$ for $e \to 0$. However, in symmetrical situations (if the
projections of $\bm{S}_1/m_1$ and $\bm{S}_2/m_2$ on the orbital plane coincide),
this derivative vanishes, due to the fact that $\tau(1,1,0) - \sigma(2,2,0) =
(P \bm{S}_1 /m_1 - P \bm{S}_2 / m_2)^2$, where $P$ is the projection operator
on the orbital plane. We can compute the value of $e^2$ for which the
derivative cancels at 2PN order, which is $e^2 = 5 x^2 [\tau(1,1,0) -
\sigma(2,2,0)]/340$. We emphasize that this effect is independent of the
particular quasi-Keplerian parametrization one chooses (see the appendix, and
in particular Eq.~\eqref{deltade2dt}).

\begin{figure}[!ht]
 \includegraphics[width=\columnwidth]{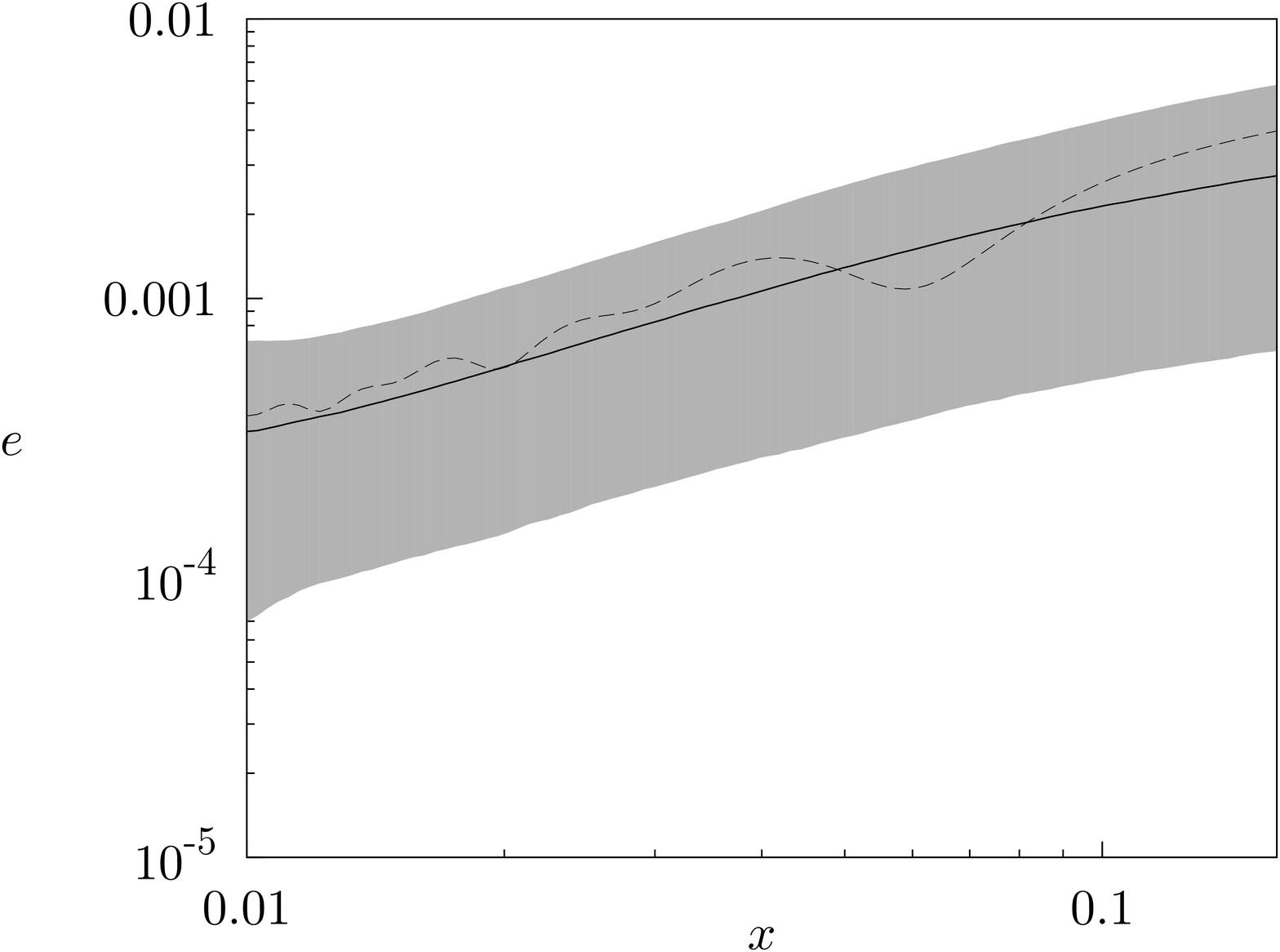}
 \includegraphics[width=\columnwidth]{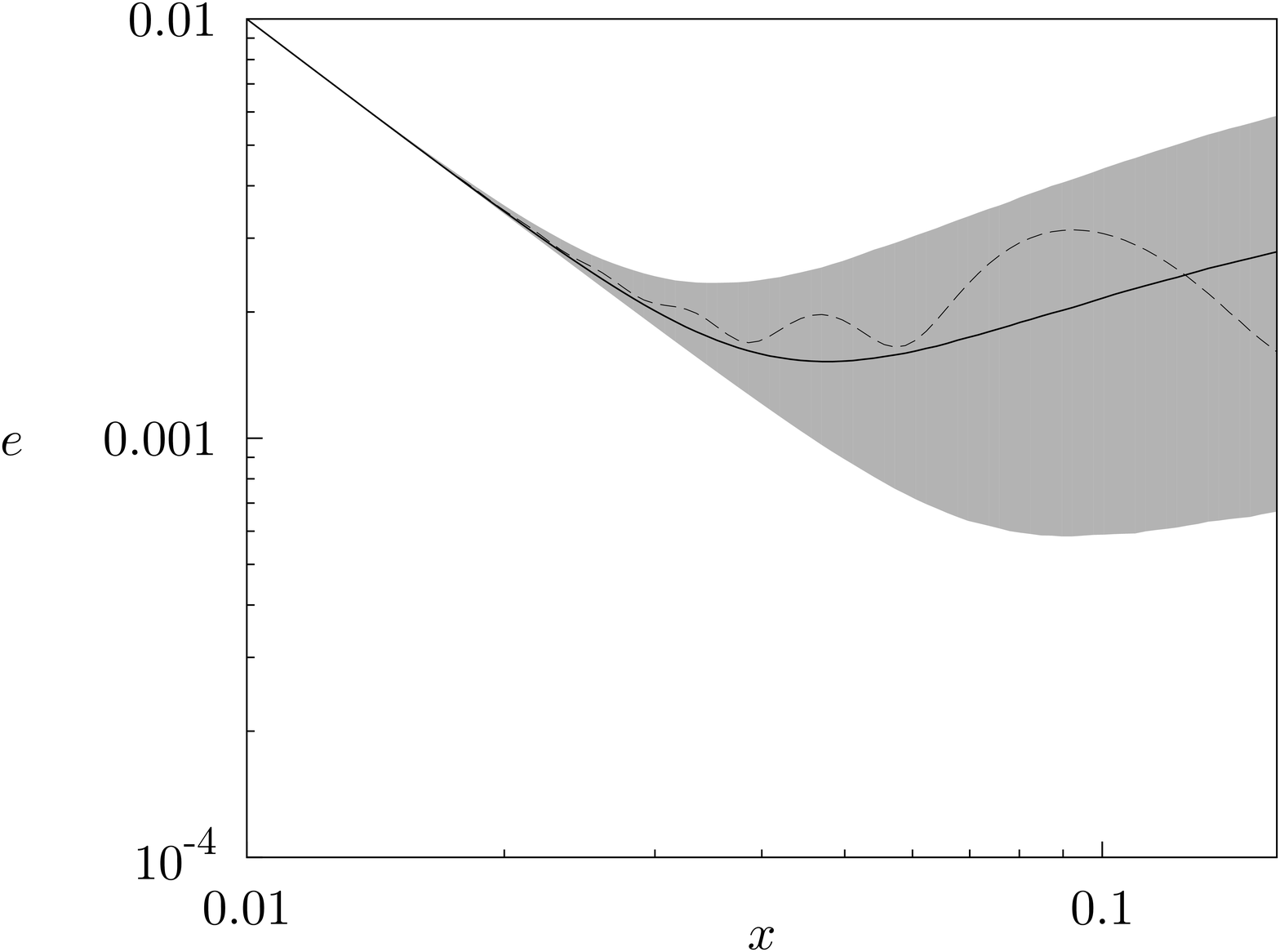}
 \caption{Evolution of the eccentricity between $x=1/100$ and $x=1/6$ with
spin-orbit and spin-spin couplings, for equal-mass binaries, at the top
starting from $e^2 = 5 x^2 [\tau(1,1,0) -
\sigma(2,2,0)]/340$, and at the bottom from $e=0.01$, with spins uniformly
distributed. In each plot, the grey region is between the $5$th and the $95$th
percentile, the solid line is the median, and the dashed line is a typical
realization.}
 \label{eofx}
\end{figure}

We plotted in Fig.~\ref{eofx} the evolution of the eccentricity between
$x=1/100$ and $x=1/6$ with
spin-orbit and spin-spin couplings, for equal-mass binaries with spins
uniformly distributed, including also the spin-independent PN corrections
computed in~\cite{dgi}, as well as spin-orbit precession~\cite{barkeroconnell}.
We see that spin-orbit precession induce a non-trivial
pattern in the evolution of the eccentricity, which could help to reduce the
errors on spin parameters in a gravitational wave measurement. We found that
the quantiles from Fig.~\ref{eofx} are very weakly
dependent on the mass ratio, whereas the amplitudes of the modulations of the
evolution of the eccentricity are strongly suppressed as the mass ratio
decreases.

\section*{Circular limit}

We define the circular limit of the quasi-Keplerian motion discussed above as
the limit $e_t \to 0$. In this limit, we also get $e_r \to 0$ and $e_\phi \to
0$. The periastron line is not well
defined, so that the equations of motion can only depend on differences of
angles.
We find (see the appendix)
\begin{align}
 r &= x^{-1} + x^{1/2} \beta(2,1) + \frac{x}{2} \gamma_1 -
\frac{x}{2} \gamma_2 \cos[2(\phi -\psi)], \\
 \frac{d\phi}{dt} &= x^{3/2} - x^{3} \beta(4,3) - \frac{x^{7/2}}{2} \left\{
3 \gamma_1 + \gamma_2 \cos[2(\phi-\psi)] \right\}.
\end{align}

We note that when one includes spin-spin couplings, the orbit can no longer
be circular in the sense that the radius depends explicitly on the angle along
the orbit, as already mentioned in~\cite{kww}. This however is not a residual
eccentricity, as the radius is
symmetric with respect to $\phi \to \phi + \pi$.

The angular frequency $d\phi/dt$ is not constant. However, we can use
Eq.~\eqref{phiofv} and define
its average along an
orbit as
\begin{align}
 \omega &= \frac{n}{2 \pi} \int_{t(v = -\pi)}^{t(v = \pi)} \frac{d\phi}{dt} dt
\nonumber\\
&=\frac{n}{2\pi} [\phi(v = \pi) - \phi(v = -\pi)] \nonumber\\
&= (-2E)^{3/2} \left[ 1 - \frac{1}{L^3} \beta(4,3) - \frac{3}{2L^4}\gamma_1
\right]. \label{omegaofEL}
\end{align}

We can thus define a new post-Newtonian parameter $z = \omega^{2/3}$. In terms
of this parameter, the constants $E$, $L$, and $x$ are
\begin{align}
 E &= - \frac{z}{2} - \frac{z^{5/2}}{3} \beta(4,3) - \frac{z^3}{2} \gamma_1, \\
 L &= z^{-1/2} - \frac{z}{6} \beta(20,15) - z^{3/2} \gamma_1, \\
 x &= z + \frac{z^{5/2}}{3} \beta(8,6) + z^3 \gamma_1.
\end{align}

Now, we can use Eqs. \eqref{omegaofEL}, \eqref{Edot}, and
\eqref{Ldot} to find
\begin{align}
 \frac{d\omega}{dt} &= \frac{96 \nu z^{11/2}}{5} \bigg[ 1 - \frac{z^{3/2}}{12}
\beta(113,75) \nonumber\\
&- \frac{z^2}{48} \sigma(247,721,0) + \frac{z^2}{96}
\tau(7,1,0) \bigg],
\end{align}
which is in agreement with what was previously computed in~\cite{kidder,mvg}.

Alternatively, if we define a circular orbit to have $de^2/dt=0$, which implies
$e^2 = 5 z^2 [\tau(1,1,0) - \sigma(2,2,0)]/340$, we get
\begin{align}
 \frac{d\omega}{dt} &= \frac{96 \nu z^{11/2}}{5} \bigg[ 1 - \frac{z^{3/2}}{12}
\beta(113,75) \nonumber\\
&- \frac{z^2}{1216} \sigma(6519,\numprint{18527},0) + \frac{z^2}{2432}
\tau(439,287,0) \bigg].
\end{align}

\section{Conclusion}

The main result of this paper is the derivation of the spin-spin effects in the
evolution of the mean motion and of the eccentricity, for binaries with an
arbitrary eccenticity. Particularly, the fact
that spin-spin couplings may induce a residual eccentricity can be important
for parameter estimation when gravitational wave observations will be made
possible. If eccentric templates allow to measure eccentricities of
$O(10^{-3}$-$10^{-4})$, the modulation induced by spin-orbit precession could
significantly improve the determination of the spins of the binary.

We also derived the equations of motion $r(t)$ and
$\phi(t)$, for black hole binaries of comparable mass with Newtonian,
spin-orbit, and spin-spin terms on eccentric orbits, and found a
family of parametrizations free of
divergencies in the circular limit $e \to 0$.

\begin{acknowledgments}
A.~K. is supported by the Swiss National Science Foundation.
\end{acknowledgments}

\appendix*

\section{A two-parameter family of quasi-Keplerian parametrizations}

We can find a two parameter family of quasi-Keplerian parametrization of the
equations of motion by taking the ansatz
\begin{align}
 r &= a \left( 1 - e_r \cos u \right) + f_{r,1} \cos(v - 2\psi)\nonumber\\
 &+
f_{r,2} \cos(2v - 2\psi),
\label{Keplereqradiusapp} \\
 \phi &= (1+k) v + f_{\phi,1} \sin(v-2\psi) + f_{\phi,2} \sin(2v-2\psi)
\nonumber\\
&+f_{\phi,3} \sin(3v-2\psi),
\label{phiofvapp}\\
 v &= 2 \arctan \left( \sqrt{\frac{1 + e_\phi}{1 - e_\phi}} \tan \frac{u}{2}
\right), \label{vofuapp}\\
 l &= n(t-t_0) = u - e_t \sin u + f_t \sin(v-2\psi),
\label{Keplereqtimeapp}
\end{align}
where $(r,\phi)$ is a polar coordinate system in the plane of motion, $n$ is the
mean motion, $u$, $v$, and $l$ are the eccentric, true, and mean anomalies, $a$
is
the
semi-major
axis, $e_t$, $e_r$, and $e_\phi$ are eccentricities, $k$ accounts for
perihelion precession, and the $f_i$ are constants.

The constants are
\begin{align}
 n &= (-2 E)^{3/2}, \\
 a &= -\frac{1}{2E} \left[ 1 - \frac{E}{L} \beta(4,2) - \frac{E}{L^2}
\gamma_1 - \lambda_1 A \frac{E}{L^2} \gamma_2 \cos2\psi \right] ,\\
 k &= -\frac{1}{L^3} \beta\left( 4, 3 \right) - \frac{3}{2L^4}
\gamma_1, \\
 e_t^2 &= A^2 + \frac{E}{L} \beta\left(8 , 6 - 2A^2 \right) \nonumber\\
 &+ 2
\frac{E}{L^2}
\left( \gamma_1 + A \lambda_2 \gamma_2 \cos2\psi \right),\\
 e_r^2 &= e_t^2 + A^2 \left[ \frac{E}{L}\beta(8,4) + 2 \frac{E}{L^2}
\left( \gamma_1 +
A \lambda_1 \gamma_2 \cos2\psi \right)
\right], \\
 e_\phi^2 &= e_t^2 + A^2 \left[ \frac{E}{L}\beta(8,8)  + 2 \frac{E}{L^2}
\left( 2\gamma_1 +
A \lambda_1 \gamma_2 \cos2\psi \right)
\right], \\
f_t &= \lambda_1 \frac{(-2 E)^{3/2}}{L} \gamma_2, \\
f_{r,1} &= - \frac{\lambda_2}{2} \frac{1}{L^2} \gamma_2, \\
f_{r,2} &= - \frac{1 + \lambda_1 A}{2} \frac{1}{L^2} \gamma_2, \\
f_{\phi,1} &= - \left[ A - \lambda_1 \left(1 + \frac{3A^2}{4}\right) -
\lambda_2 \right] \frac{1}{L^4} \gamma_2, \\
f_{\phi,2} &= -\frac{1 - 4 A \lambda_1 - A \lambda_2}{4} \frac{1}{L^4} \gamma_2,
\\
f_{\phi,3} &= \frac{\lambda_1 A^2}{4} \frac{1}{L^4} \gamma_2,
\end{align}
where $\lambda_1$ and $\lambda_2$ are arbitrary functions of $A$.

The Keresztes-Mik\'oczi-Gergely (KMG) quasi-Keplerian
parametrization of the orbit~\cite{kmg} is obtained by imposing $f_{r,i} = 0$,
which leads to $\lambda_2 = 0$ and $\lambda_1 = -1/A$.

To get Eqs.~\eqref{Keplereqradius} to~\eqref{Keplereqtime}, we required that the
values of
the semi-major axis and of the eccentricities
should not depend on the position of the periastron line in the orbital
plane. This implies $\lambda_1 = \lambda_2 = 0$.

We can express $E$ and $L$ as functions of the post-Newtonian parameter $x =
n^{2/3}$ and of the eccentricity $e = e_t$. We get
\begin{align}
 E &= - \frac{x}{2}, \\
 L &= \frac{\sqrt{1 - e^2}}{x^{1/2}} \Bigg[ 1 - \frac{x^{3/2} \beta \left(4, 3-
e^2 \right)}{2\left(1-e^2\right)^{3/2}} \nonumber\\
&-
\frac{x^2}{2\left(1-e^2\right)^2} \left( \gamma_1 + e \lambda_2 \gamma_2
\cos2\psi\right) \Bigg].
\end{align}

The constants in the equations of motion then become
\begin{align}
 a &= x^{-1} \bigg[ 1 + \frac{x^{3/2} \beta(2,1)}{\sqrt{1-e^2}} \nonumber\\
 &+
\frac{x^2}{2\left(1-e^2\right)} \left( \gamma_1 + e \lambda_1 \gamma_2 \cos2\psi
\right) \bigg] , \\
 k &= -\frac{x^{3/2} \beta\left(4,3\right)}{\left(1-e^2\right)^{3/2}} -
\frac{3x^2
\gamma_1}{2\left(1-e^2\right)^2} ,\\
 e_r &= e \bigg[ 1 - \frac{x^{3/2} \beta \left(2, 1
\right)}{\sqrt{1-e^2}} \nonumber\\
 &- \frac{x^2}{2\left(1-e^2\right)} \left(\gamma_1 + e
\lambda_1 \gamma_2 \cos2\psi \right)\bigg] ,\\
 e_\phi &= e \bigg[ 1 - \frac{x^{3/2} \beta \left(2, 2
\right)}{\sqrt{1-e^2}} \nonumber\\
 &- \frac{x^2}{2\left(1-e^2\right)} \left(2 \gamma_1 + e
\lambda_1 \gamma_2 \cos2\psi \right)\bigg] ,\\
f_t &= \frac{x^2}{\sqrt{1-e^2}} \lambda_1 \gamma_2, \\
f_{r,1} &= - \frac{x}{2\left(1 - e^2\right)} \lambda_2 \gamma_2, \\
f_{r,2} &= - \frac{x}{2\left(1 - e^2\right)} (1 + e \lambda_1) \gamma_2,\\
f_{\phi,1} &= - \frac{x^2}{\left(1-e^2\right)^2} \left[e - \lambda_1 \left(1 +
\frac{3e^2}{4} \right) - \lambda_2\right] \gamma_2 , \\
f_{\phi,2} &= - \frac{x^2}{\left(1-e^2\right)^2} \left(
\frac{1 - 4 e \lambda_1 - e\lambda_2}{4} \right) \gamma_2 , \\
f_{\phi,3} &= \frac{x^2}{\left(1-e^2\right)^2} \frac{e^2 \lambda_1}{4}
\gamma_2.
\end{align}

One can see that for the solution to be free of divergencies in the zero
eccentricity limit, $\lambda_1$ and $\lambda_2$ must be regular as $e \to 0$,
which implies, at this PN order, that they must be regular functions of
$A$. It is not the case for the KMG parametrization, but this is
in fact a coordinate singularity, since, as we will see later, the equations of
motion have a well-defined zero-eccentricity limit.

The effect of $\lambda_1$ and $\lambda_2$ on the evolution of the eccentricity
with respect to Eq.~\eqref{edot} is
\begin{equation}
 \delta \frac{de^2}{dt} = \frac{\nu x^6 \lambda_2}{15 \left(1-e^2\right)^{9/2}}
\sigma\left( 0 , 0, 688 e + 2139 e^3 + 148 e^5 \right), \label{deltade2dt}
\end{equation}
which, as long as the parametrization is regular, does not affect the residual
eccentricity we found in this paper.

\section*{Circular limit}

Let us now compute the equations of motion $r(\phi)$ and $d\phi/dt$.
From Eq.~\eqref{phiofvapp}, we get
\begin{align}
 v &= (1-k) \phi - f_{\phi,1} \sin(\phi-2\psi) - f_{\phi,2} \sin(2\phi-2\psi)
\nonumber\\
&-f_{\phi,3} \sin(3\phi-2\psi).
\end{align}

Then, together with Eq.~\eqref{vofuapp}, Eq.~\eqref{Keplereqradiusapp}
becomes, at leading order in $e$,
\begin{align}
 r &= x^{-1} + x^{1/2} \beta(2,1) + \frac{x}{2} \gamma_1 - \frac{x}{2}
\gamma_2 \big\{ \cos[2(\phi-\psi)]
 \nonumber\\
& + \left(\lambda_2 + e^2 \lambda_1 \right) \cos(\phi - 2\psi) - e^2 \lambda_1
\cos(\phi + 2\psi) \big\}
 ,
\end{align}
and we can compute from Eq.~\eqref{Keplereqtimeapp}, at leading order in $e$,
\begin{align}
 \frac{d\phi}{dt} &= x^{3/2} - x^3 \beta(4,3) - \frac{3 x^{7/2}}{2} \gamma_1  -
\frac{x^{7/2}}{2} \gamma_2 \cos[2(\phi -\psi)] 
\nonumber\\
& + \gamma_2 x^{7/2} \big[\left(\lambda_2 + 2 e^2 \lambda_1 \right)
\cos(\phi - 2\psi) \nonumber\\
&- 2 e^2 \lambda_1 \cos(\phi +
2\psi)\big] .
\end{align}

One can see that the use of the KMG
parametrization does not change the equations of motion in the zero eccentricity
limit with respect to the one obtained from Eqs.~\eqref{Keplereqradius}
to~\eqref{Keplereqtime}. Furthermore, as the periastron line is not
well-defined in the circular limit, we have to impose that $\lambda_2 \to 0$ as
$e \to 0$
(which is equivalent to $\lambda_2 \sim O(A)$), so that the equations of motion
in the circular limit are independent of the choice of an arbitrary periastron
line. From Eq.~\eqref{phiofvapp}, we see that such an arbitrary choice induces a
non-zero value for $\phi(v=0)$ of 2PN order. However, the differences induced in
the equations of motion are subsequently of 4PN order, far beyond the limit of
their
validity.

\end{document}